# FaceAtlasAR: Atlas of Facial Acupuncture Points in Augmented Reality

Menghe Zhang[1], Jürgen P. Schulze[2], and Dong Zhang[3]

[1,2]Department of Computer Science, University of California, San Diego
[1]mez071@ucsd.edu, [2]jschulze@ucsd.edu
[3]Qilu University of Technology, Shandong, China
jnzd156@163.com

## ABSTRACT

*Acupuncture is a technique in which practitioners stimulate specific points on the body. Those points, called acupuncture points (or acupoints), anatomically define areas on the skin relative to specific landmarks on the body. However, mapping the acupoints to individuals could be challenging for inexperienced acupuncturists. In this project, we proposed a system to localize and visualize facial acupoints for individuals in an augmented reality (AR) context. This system combines a face alignment model and a hair segmentation model to provide dense reference points for acupoints localization in real-time (60FPS). The localization process takes the proportional bone (B-cun or skeletal) measurement method, which is commonly operated by specialists; however, in the real practice, operators sometimes find it inaccurate due to the skill-related error. With this system, users, even without any skills, can locate the facial acupoints as a part of the self-training or self-treatment process.*

## 1. INTRODUCTION

Acupuncture [1] is a form of alternative medicine and a key component of traditional Chinese medicine (TCM). Based on the symptoms, acupuncturists stimulate specific anatomic sites commonly by needling, massaging, or heat therapy. Scientific studies have proved that acupuncture may help ease types of pain that are often chronic such as low-back pain, neck pain, and osteoarthritis/knee pain. The acupoints on the face can help with a variety of conditions both on and off the face, such as jaw tension, headaches, anxiety, and stomach conditions.

However, acupuncture practice relies on experienced acupuncturists to locate the acupoints from body acupuncture maps. Individuals, who want to help themselves relieve symptoms with acupoints stimulation, usually find it confusing to localize the targets by natural language description or pictures of a standard model.

With the help of augmented reality, we designed a system to display facial acupoints on the top of the user's face to accurately view and localize facial acupuncture points. Specifically, we employ a deep learning model to annotate 3D landmarks and on a user's face together with hair segmentation in real-time to gather reference points for acupuncture points localization. We then align the reference point with acupuncture points defined by proportional bone measurement method. The whole process is implemented via MediaPipe [2], a framework for building cross-platform machine learning solutions. Our system works perfectly on phones with a front camera, without any extra hardware, users could pinpoint and interact with the target acupoints. There are three benefits to this application:

- It takes a conventional localization method while originally defines a scheme to transform the natural language descriptions to mathematical logic expressions.
- It adopts MediaPipe framework to run across platforms in real-time.
- It adapts to different head poses to be robust for users to locate acupoints in different regions.

With FaceAtlasAR, people who have little or no experience in localizing facial acupuncture points can use. Potential use scenarios for our applications are varied, for example, acupuncture education, communication, and self-healing.

## 2. RELATED WORK

### 2.1. Acupuncture Points Localization

Existing works of acupuncture training applications on AR devices are limited. In 2015, H. Jiang et al. [3] proposed the first acupuncture training application, Acu Glass, on a head-mount display device (HMD) based on Google Glass. They generated the frontal face acupoints based on the height and the width of the input face, plus the distance between the eyes. However, their face landmarks for reference are too limited to adapt to different people and different poses of the face. Other acupoints localization methods like Chen et al. [4][5] fit a 3D Morphable Face Model (3DMM) [6] to a 2D image and combine facial landmarks and image deformation to estimate acupoints. Although 3DMM is a powerful tool to build polygonal mesh, the range of possible predictions is limited by the linear manifold spanned by the PCA basis, which is in turn determined by the diversity of the set of faces captured for the model [7]. Therefore, manual annotation on a standard 3D model may not correctly fit all kinds of people. Moreover, acupoints are officially defined relative to landmarks, while the deformation process does not guarantee the relativeness.

As for the localization methods in practice, Godson and Wardle [8] screened 771 studies and summarized the methods as Directional(F-cun) method, Proportional method, palpation for tenderness, electronic point detectors, and anatomical locations. Usually, more accurate approaches are the next steps to less accurate ones and require extra hardware. For example, one can roughly locate a target by the directional method and then use electronic point detectors to measure the electrical resistance of the skin. There is research related to this topic and acupoint probing devices available in the market.

### 2.2. Face Alignment

Face alignment is a computer vision technology for identifying the geometric structure of human faces in digital images. Bulat and Tzimiropoulos [9] reviewed 2D and 3D face alignment and landmark localization. Existing 2D and 3D datasets annotate a limited set of landmarks. For example, 300-W [10], the most widely used in-the-wild dataset for 2D alignment, containing LFPW [11], HELEN [12], AFW [13], and iBUG [10], annotates only 68 landmarks per face. These landmarks either have distinct semantics of their own or participate in meaningful facial contours. Works based on them are not suitable for our cases. We finally adopted the work of Kartynnik et al. [7], which estimates 3D mesh with 468 vertices in real-time. The vertices are selected manually according to expressive AR effects, thus well suitable for our requirements.

## 3. SYSTEM OVERVIEW

Our facial localization solution utilizes the MediaPipe machine learning pipeline consisting of an off-line stage and an on-line stage (Fig.1).

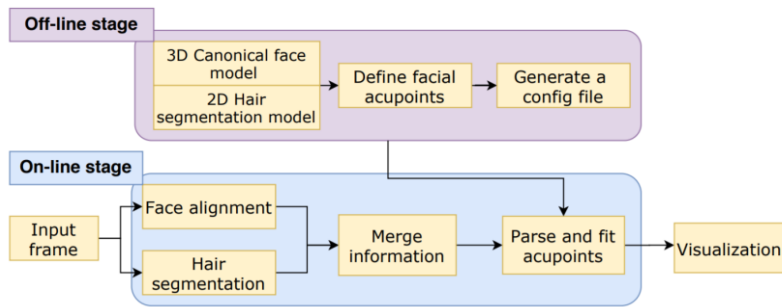

Figure 1: Workflow of the proposed system

During the off-line stage:

- **Model selection**: Based on the definition of the referenced points on a face, we choose to blend a pre-trained face alignment model and a pre-trained hair segmentation model. This is because a single model cannot cover the whole set of target regions, while different parts of the regions are in the different reference systems.

- **Acupoints localization**: We firstly localize facial anatomical landmarks, which are designated by the National Standard of the People's Republic of China. We then use the proportional bone (B-cun or skeletal) measurement method to locate all acupoints on the face.

- **Data file generation**: The file contains all the information needed for each acupoint/reference point, including name, region, relative location towards a landmark or a reference point. The file is readable for non-technical users, for example, acupuncturists, to correct less accurate acupoint descriptions by natural language.

Then at the on-line stage, the system gets face landmarks together with hair segmentation and merges those results into acupoints generator. The generator gives out the requested acupoints on this frame based on the prior knowledge and then draw on the input face.

We fit the whole process into MediaPipe's perception pipeline as a graph of modular components. Each component, called Calculator, solves an individual task like model inference, data transformation, or annotation. We will talk about the implementation details in the next section. We show the graph for our FaceAtlasAR in Figure 2.

The graph consists of two subgraphs: one for face alignment (FaceLandmarkFrontGpu) and the other for hair segmentation (HairSegmentationGpu). From the graph, we see the FlowLimiter guards the whole pipeline process. By connecting the output of the final image to the FlowLimiter with a backward edge, the FlowLimiter keeps track of how many timestamps are currently being processed. The system will down-sample and transform the original image before fusing the input to machine learning models but will draw the results onto the original frames. The next section will show the implementation of each module in detail.

## 4. IMPLEMENTATION

### 4.1 Face Alignment

We adopted a pre-trained TFLite model [7] to infer an approximate 3D mesh representation of a human face.

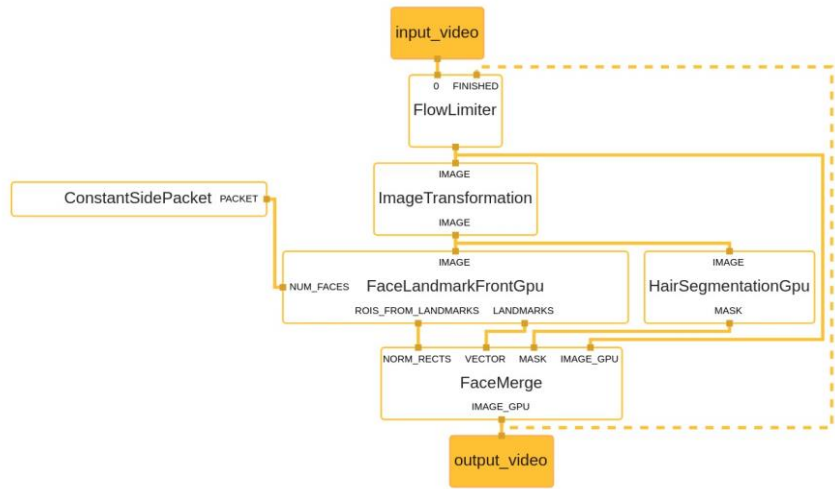

Figure 2: The graph of FaceAtlasAR

This process comprises of majorly three steps:

- **Face detection**: The whole frame is first processed by a lightweight face detector to get the face bounding box and several landmarks, thus, to get the rotation matrix of the face. This step only runs until the system finds a face to track or when the system loses tracking.

- **Image transformation**: The image is then cropped by the bounding box and resized to fit into the next step. After this step, the target region is centered and aligned.

- **Face landmarks generation**: The pre-trained model produces a vector of 3D landmark coordinates, which subsequently gets mapped back into the original image coordinate system.

Then from a canonical face mesh model, we extract those vertices with semantic meaning as the reference points (Fig.3).

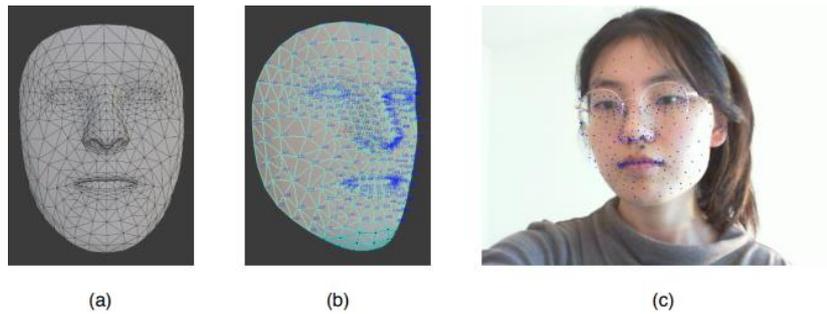

Figure 3: The generated mesh topology(a), its vertices with index(b), and viewed in AR(c)

### 4.2. Hair Segmentation

Since the center of the frontal hairline is a critical facial anatomical landmark according to the national standard, we adopted a pre-trained model [14] to get the hair segmentation mask. Similar to the face alignment process, the previously generated mask can be fed back to help accelerate the process. Specifically, the mask from the previous round of inference will be embedded as the alpha channel of the current input image (Fig.4).

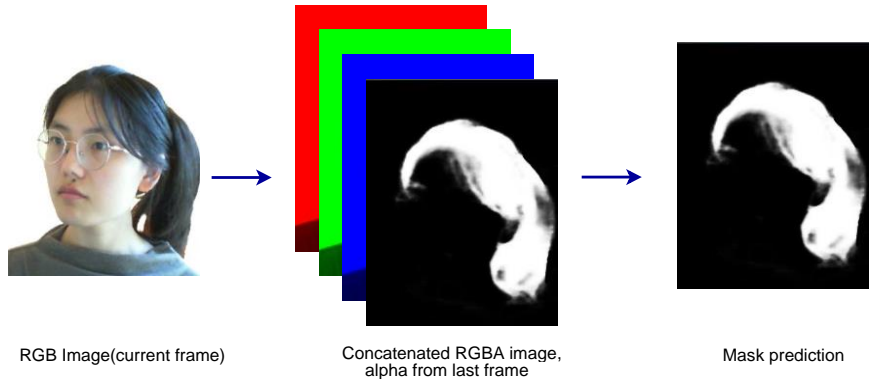

Figure 4: Hair segmentation module

### 4.3. Facial Acupoints Localization

Given the hair mask together with face landmarks, we now could locate facial acupoints based on the B-cun method. This refers to the method of measuring the length and width of each part of the body with the body surface condyles as the main landmark and determining the position of acupoints. Then, a unit "cun" is the length between the set two bone nodes divided into certain equal parts as the basis for setting acupoints. The facial acupoints bank on the unit cun's definition of the head as shown in Figure 5. To start with, we locate three facial anatomical landmarks in consonance with the national standard. In order to differentiate these three points to acupoints, we group them in the channel named RHD:

- **RHD1, Yintang:** The midway between the medial ends of the eyebrows.
- **RHD2, Middle of anterior hairline:** Intersection of anterior hairline and anterior midline.
- **RHD3, Pupils**

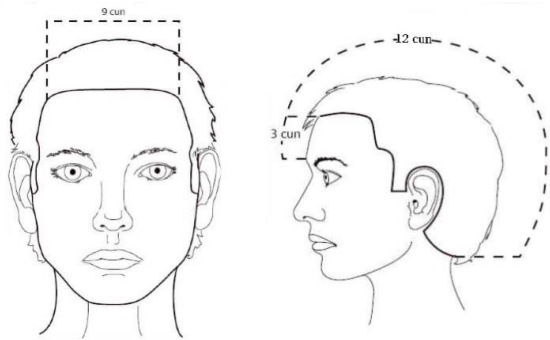

(a) B-cun on the font face  (b) B-cun on the side face

Figure 5: B-cun on the head from National Standard of the People's Republic of China, Acupoints [15]. Pictures from [16].

Table 1: Information of an example acupoint Sibai (ST2) in the data file.

| Channel Name | ID | NameE | Region |
|---|---|---|---|
| ST | 2 | Sibai | eye |
| FaceMeshX | FaceMeshY | IsSymmetry | Comments |
| GetX(RHD3) | GetY(ST1)+0.5*U | TRUE | - |

From Fig.5b, the distance from Yintang to Middle of anterior hairline, $d = (p_{RHD1} - p_{RHD2})$, decides the unit cun as $uc = d/3$.

On that occasion, like RHD points definition, we could locate all facial acupoints. Table1 shows an example of what information we keep for each acupoint. All points' information finally makes up to a data file.

There is one more aspect we want to specify, the channel name, which refers to a unique meridian channel. The meridian system[17] is a concept in TCM about a path through which the life-energy is known as "qi" flows. There are 12 standard and 8 extraordinary meridians, while acupoints are the chosen sites on the meridian system. We group the acupoints on the same meridian channel and connect them by the flow. For example, the previously stated acupoint, Sibai, belongs to the ST(Foot's Yang Supreme Stomach Meridian) channel. Acupoints on the ST channel and their flow are illustrated in the Figure 6.

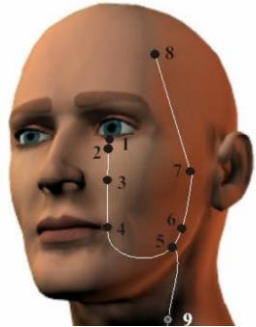

Figure 6: Illustration of ST channel on the head. Picture from [18]

## 5. RESULTS

We achieve real-time performance on both desktop and mobile devices by designing the pipeline properly. We show the final application below and compare the performance on different platforms.

### 5.1. Android Application

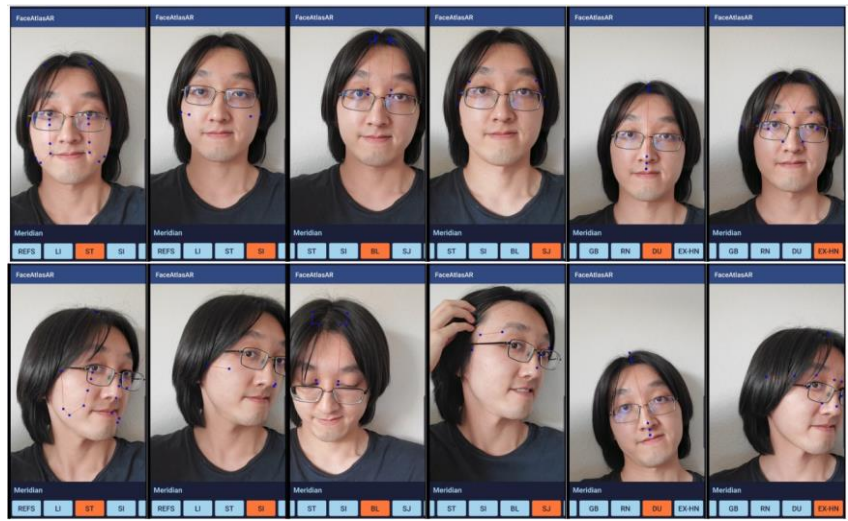

Figure 7: Android application screenshots for displaying acupoints grouped by meridian system in different poses

Figure 7 presents the screenshots from our FaceAtlasAR android app. Here we show the visualization of acupoints grouped by meridian channels in different poses. Thanks to the robustness of face alignment towards occlusion, users would not find problems pointing or pressing a target acupoint (Fig. 8).

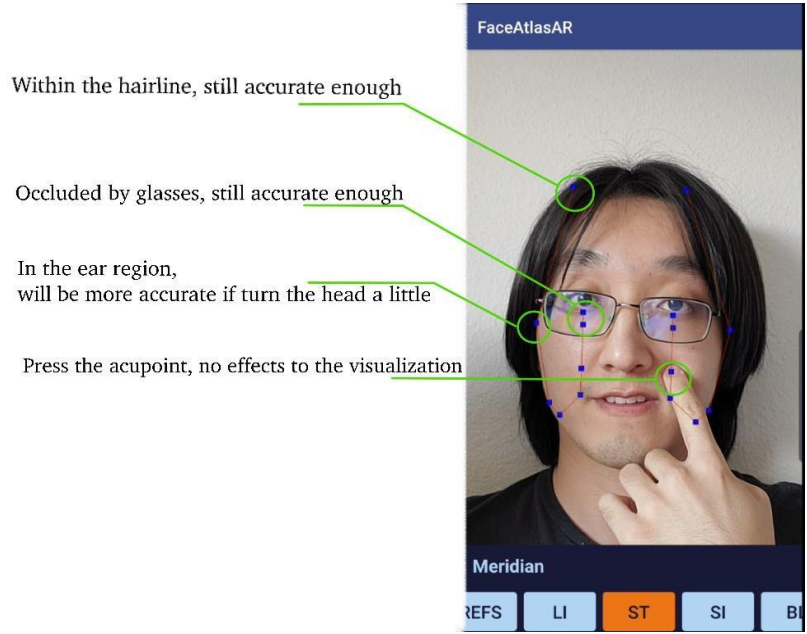

Figure 8: User presses an acupoint on ST meridian channel

### 5.2. Accuracy

We setup an experiment on Samsung S10($2280 \times 1080$) to valid the accuracy of our FaceAtlasAR to localize 4 reference points and 69 acupuncture points on face by comparing the mean pixel errors between ground truth positions and the localization results. Initially, we assign all the target positions into three groups by the localization complexity as shown in the Table 2.

Table 2: Data file parsing performance

| Quantity | Directly from face alignment results | One-time proportional localization | Multiple-times proportional localization |
|---|---|---|---|
| Reference points | 3 | 1 | 0 |
| Acupoints | 38 | 16 | 15 |

We then investigate the mean pixel errors in 3 groups. For each localization points, we measure the pixel errors in 4 different poses: frontal face (0º), pitch (X-axis +10º), roll (Y-axis +10º), yaw (Z-axis +10º) to get the mean value. The results are shown in Figure 9.

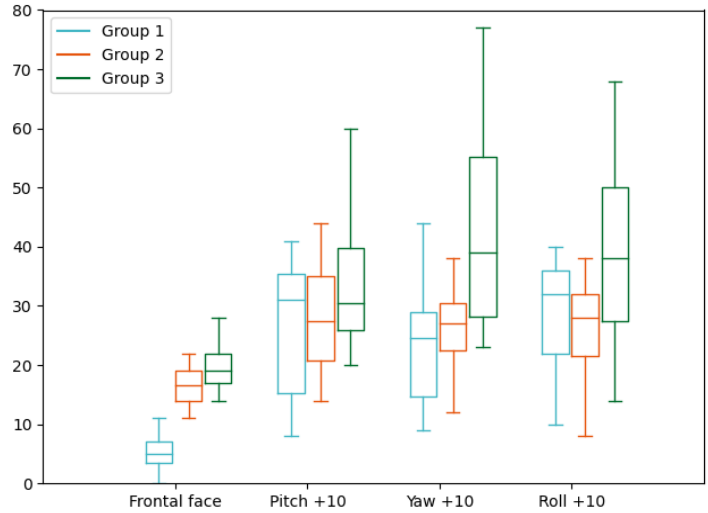

Figure 9: Mean pixel errors of localization

From the results we found that some of the localizations are less precise than others. Especially, the system performs worse on group 2 than group 1 and group 3 than group 2. Accordingly, multiple times of proportional calculations that rely on the cun system add inaccuracy to the results. The system tolerates to different angles to some extent. However, a target point will be more well localized when it is point straight to the camera without any angles.

### 5.3. Performance

We evaluate the performance of our pipeline in three major components: face alignment, hair segmentation, and acupoints generation. Since the application runs on multiple platforms, we compare its performance on a desktop with Nvidia GeForce RTX 2070 SUPER and on a Samsung S10. The input images for two TFLite models are both in full size 512×512. We see that in this case, Samsung S10 still runs in full frame rate (60FPS). More detailed comparison is as in Table 3. We also evaluate the performance of data file parsing (Table 2), which only runs once at the setup stage.

Table 3: Data file parsing performance

| Device/Time, ms | File parser |
| --- | --- |
| Desktop | 0.102 |
| Samsung S10 | 0.279 |

Table 4: Application performance on a desktop and a mobile device

| Device/Time(ms) | Hair segmentation | Face alignment | Generation | Overall |
| --- | --- | --- | --- | --- |
| Desktop | 1.188 | 3.376 | 0.135 | 10.56 |
| Samsung S10 | 50.279 | 14.673 | 0.512 | 84.758 |

### 6. DISCUSSION

From the results we could see that:

- Our system can properly display the requested acupoints on selected meridian channels.
- The system tolerates movements very well, while endures tilt and rotation to some degree.
- The benefit from head rotation is that: when the acupoints are hidden in one view, they will be visible and more accurate in another one. For example, the frontal face hides acupoints in the ear region; thus, to view them around the left ear properly, the user needs to turn his/her head so that the left ear faces the camera.

Our next step is to improve the face alignment performance on the side of the head. There are dozens of acupuncture points around the ears that represent specific domain and functions of the body. However, most face alignment jobs only require a small set of landmarks. Even though the model we adopted can estimate 3D mesh with 468 vertices, it still neglects both sides of the head and loses some accuracy at the cheeks and chins. Therefore, we could only roughly estimate those acupoints' positions based on the proportional relations to other landmarks on face, which is less meticulous.

## 7. CONCLUSION

In this paper, we proposed FaceAtlasAR, an end-to-end facial acupoints tracking solution that achieves real-time performance on mobile devices. Our pipeline integrates a face alignment model with a hair segmentation model. The high accuracy of the estimation and the robustness of the system empower users with little experience in acupuncture to interact with facial acupoints. Future work comes to improving the accuracy even further in the ear region since the face alignment only gives a little information towards the face edge near the ear region. Also, 3D interaction should be considered for users to gain a more immersive experience. For example, we could track users' hands/fingers while they are interacting with a target acupoint.

**Authors**

Menghe Zhang

Graduate student, University of California, San Diego

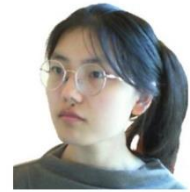

Jürgen P. Schulze

Associate Research Scientist, Qualcomm Institute, UCSD

Associate Adjunct Professor, Department of Computer Science, UCSD

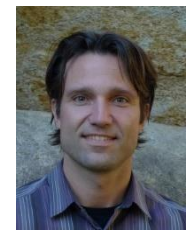

Dong Zhang

Research Scientist, Qilu University of Technology

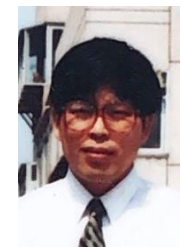